\documentclass{llncs}
\usepackage{makeidx}  
\usepackage{graphicx} 
\usepackage{subeqnar} 
\usepackage{multicol} 
\usepackage{cropmark} 

\usepackage{amsmath}
\usepackage{amssymb}
\usepackage{amscd}
\usepackage{latexsym}
\usepackage{epsfig}

\newtheorem{defi}{Definition}
\newtheorem{thm}[defi]{Theorem}

\newtheorem{exempel}[defi]{Example}

\newlength{\blank}

\date{}

\begin{document}

\pagestyle{plain}  
\pagenumbering{arabic}

\title{The Complexity of Probabilistic versus Quantum Finite Automata}


\titlerunning{The Complexity of Probabilistic versus Quantum Finite Automata}

\author{Gatis Midrij\= anis
\thanks{Research supported by Grant No.01.0354 from the
Latvian Council of Science, and  Contract IST-1999-11234 (QAIP)
from the European Commission.}}
\authorrunning{Gatis Midrij\= anis }

\institute{University of Latvia, Rai\c na bulv\= aris 19, Riga,
Latvia. Email: \texttt{mgatis@inbox.lv}. Fax:
\texttt{+371-7820153}.}


\maketitle

\begin{abstract}
 We present a language $L_n$ which is recognizable by
  a probabilistic finite automaton (PFA) with probability $1 -
  \epsilon$ for all $\epsilon > 0$ with $O(log^2n)$ states, with
  a deterministic finite automaton (DFA) with $O(n)$ states, but a
  quantum finite automaton (QFA) needs at least $2^{\Omega(n/ \log n)}$ states.
\end{abstract}

\section{Introduction}
\label{sec:intro} \label{sec:defi} \par

\par
A PFA is generalization of DFA.  Many authors have tried to find
out (~\cite{a:f}, ~\cite{freiv}, ~\cite{a}, ~\cite{ras} a. o.) the
size advantages of PFA over DFA. On the other side it is known
(~\cite{a:n:t:v}, ~\cite{a:f}) that  the size of reversible finite
automata (RFA) and  the size of QFA exceed the size of the
corresponding DFA almost exponentially for some regular languages
(i.e. for languages recognizable by  DFA). And so A. Ambainis, A.
Nayak, A. Ta-Shma, U. Vazirani ~\cite{a:n:t:v} wrote:
\begin{quote}
  {\it
  Another open problem involves the blow up in size while
  simulating a 1-way PFA by a 1-way QFA. The only known way for doing
  this is by simulating the PFA by a 1-way DFA and then simulating
  the DFA by a QFA. Both simulating a PFA by a DFA (~\cite{a}, ~\cite{freiv},
  ~\cite{ra}) and simulating DFA by a QFA (this paper) can involve
  exponential or nearly exponential increase in size. This means
  that the straightforward simulation of a probabilistic automaton
  by a QFA (described above) could result in a doubly-exponential
  increase in size. However, we do not known of any examples where
  both transforming a PFA into a DFA and transforming a DFA into a
  QFA cause big increases of size. Better simulations of PFA by
  QFAs may well be possible.}
\end{quote}
We will solve this problem.

\section{Definitions and known results}
\label{sec:known} \par We use the definition of 1-way QFA (further in text simply
QFA) as in~\cite{a:f} and ~\cite{a:n:t:v}. This model was first introduced in
\cite{kondacs:watrous} and is not the most general one, but is easy to implement and
deal with. A quantum finite automaton has a finite set of basis states $Q$, which
consists of tree parts: accepting states ($Q_{acc}$), rejecting states ($Q_{rej}$)
and non-halting states ($Q_{non}$). One of the states, $q_{ini}$, is distinguished
as the starting state.
\par Inputs to a QFA are words over a finite alphabet $\Sigma$. We
shall also use the symbols $\o$ and $\$$ that do not belong to
$\Sigma$ to denote the left and the right end marker,
respectively. The set $\Gamma = \Sigma \ \cup \ \{\o, \$\}$
denotes the working alphabet of the QFA. For each symbol $\sigma
\in \Gamma$, a QFA has a corresponding unitary transformation
$U_{\sigma}$ on the space $\mathbb{C}^Q$.
\par At any time, the state of a QFA is a superposition of basis
states in $Q$. The computation starts in the superposition
$|q_{ini}\rangle$. Then the transformations corresponding to the
left end marker $\o$, the letters of the input word $x$ and the
right end marker $\$$ are applied in succession to the state of
the automaton, unless a transformation results in acceptance or
rejection of the input. A transformation consists of two steps:
\begin{enumerate}
    \item First, $U_{\sigma}$ is applied to $|\psi\rangle$, the current
    state of the automaton, to obtain the new state $|\psi'\rangle$.
    \item Then, $|\psi'\rangle$ is measured with respect to the
    observable $E_{acc}\ \oplus \ E_{rej}\ \oplus \ E_{non}$,
    where $E_{acc} = span\{|q\rangle \ | q \in Q_{acc}\}$,
    $E_{rej} = span\{|q\rangle \ | q \in Q_{rej}\}$,
    $E_{non} = span\{|q\rangle \ | q \in Q_{non}\}$. The
    probability of observing $E_i$ is equal to the squared norm of
    the projection of $|\psi'\rangle$ onto $E_i$. On measurement,
    the state of the automaton "collapses" to the projection onto
    the space observed, i.e., becomes equal to the projection,
    suitably normalized to a unit superposition. If we observe
    $E_{acc}$ (or $E_{rej}$), the input is accepted (or rejected).
    Otherwise, the computation continues, and the next
    transformation, if any, is applied.
\end{enumerate}
A QFA is said to $accept$ (or $recognize$) a language $L$ with
probability $p > \frac{1}{2}$ if it accepts every word in $L$ with
probability at least $p$, and rejects every word not in $L$ with
probability at least $p$.
\par A RFA is a QFA with elements only
$0$ and $1$ in the matrices. A PFA is the same as a QFA but only
instead of unitary matrices it has stochastic ones. A DFA is a PFA
with only $0$ and $1$ in the matrices.
\par The $size$ of a finite automaton is defined as the number of
(basis) states in it. More exact definitions one can find, for
example, in~\cite{a:f}.
\par In~\cite{a:f} there was given a
language $L^\times_n$ consisting of one word $a^n$ in a
single-letter alphabet and it was proved:
\begin{thm}\par
  \label{thm:pd} \par
    \begin{enumerate}\par
      \item Any deterministic automaton that recognizes $L^\times_n$, has at least n states.
      \item For any $\epsilon > 0$, there is a probabilistic automaton with $O(\log^2n)$ states
        recognizing $L^\times_n$ with probability $1 - \epsilon$.
    \end{enumerate}
\end{thm}
{\it Sketch of Proof.} The first part is evident. To prove the
second part, Freivalds ~\cite{freiv} used the following
construction. $O(\frac{\log n}{ \log \log n})$ different primes
are employed and $O(\log n)$ states are used for every employed
prime. At first, the automaton randomly chooses a prime p, and the
the remainder modulo p of the length of input word is found and
compared with the standard. Additionally, once in every p steps a
transition to a rejecting state is made with a "small" probability
$\frac{const p}{n}$. The number of used primes suffices to assert
that, for every input of length less than n, most of primes p give
remainders different from the remainder of n modulo p. The "small"
probability is chosen to have the rejection high enough for every
input length N such both $N \neq n$ and $\epsilon$-fraction of all
the primes used have the same remainders mod p as n. \qed

In ~\cite{a:n:t:v} was definition and theorem:
\begin{defi}
    \label{defi:enc}
    $f : \{0, 1\}^m \times R \longmapsto \mathbb{C}^{2^n}$ serially encodes m classical bits
    into n qubits with p success, if for any $i \in [1..n]$ and $b_{[i+1,n]} = b_{i+1}...b_n \in \{0, 1\}^{n-i}$,
    there is a measurement $\Theta_{i,b_{[i+1,n]}}$ that returns $0$ or $1$ and has property that \par
    $\forall b \in \{0, 1\}^m : Prob(\Theta_{i, b_{[i+1,n]}} | f(b,r)\rangle = b_i) \geq p$. \par
\end{defi}
\begin{thm}
    \label{thm:encod}
    Any quantum serial encoding of m bits into n qubits with
    constant success probability $p > 1/2$ has $n \geq \Omega(\frac{m}{\log m})$.
\end{thm}
And also in ~\cite{a:n:t:v} there was defined an r-restricted
1-way QFA for a language L as a 1-way QFA that recognizes the
language with probability $p > 1/2$, and which halts with non-zero
probability before seeing the right end marker only after it has
read r letters of the input. \par The following theorem was
proved:
\begin{thm}
    \label{thm:rr}
    Let M be a 1-way QFA with S states recognizing a language L
    with probability $p$. Then there is an r-restricted 1-way QFA
    $M^{'}$ with $O(rS)$ states that recognizes L with probability $p$.
\end{thm}

\section{Results} \label{sec:res} \par

\par
One of the components of the proof of Theorem \ref{thm:pr} below
is the following lemma:

\begin{lemma}\par
  Language
  \label{lemma:div0} $${L_1}={\{\omega \in \{0, 1\}^* : \exists x, y \in \{0, 1\}^* : \omega = x00y\}}$$\par
  is recognizable by a DFA.
\end{lemma}
{\it Sketch of Proof.} The automaton has five states: $q_0$,
$q_1$, $q_2$, $q_{acc}$ and $q_{rej}$. Values of the transition
function between states are: f($q_0$, 0) = $q_1$, f($q_0$, 1) =
$q_0$, f($q_1$, 0) = $q_2$, , f($q_1$, 1) = $q_0$, f($q_2$, 0) =
$q_2$, f($q_2$, 1) = $q_2$, f($q_0$, \$) = $q_{rej}$, f($q_1$, \$)
= $q_{rej}$, f($q_2$, \$) = $q_{acc}$. \qed\par

\begin{thm}\par
  \label{thm:pr}
  For all $k \geq 1$, n = 2*k, we define language \par
  $${L_n}={\{\omega \in \{0, 1\}^n : \exists x, y \in \{0, 1\}^* : \omega = x00y \}}.$$\par
    \begin{enumerate}
      \item[0.] There is a RFA (so also a QFA, a PFA and a DFA) that recognize $L_n$.
      \item Any RFA that recognizes $L_n$, has at least $2^{O(n)}$ states.
      \item Any QFA that recognizes $L_n$ with probability $p > 1/2$, has at least $2^{\Omega(\frac{n}{\log n})}$ states.
      \item Any DFA that recognizes $L_n$, has at least $O(n)$ states.
      \item For any $\epsilon > 0$, there is a PFA with $O(\log^2n)$ states
        recognizing $L_n$ with probability $1 - \epsilon$.
    \end{enumerate}
\end{thm}
{\it Proof.}
\par Zero part follows from fact that all finite languages are
 recognizable by some RFA and $L_n$ is finite language.
\par First part: We give to automaton word
 $a_11a_21a_31a_41a_51a_61...a_k1$, where $a_i \in \{0, 1\}$.
 It is obvious that then automaton cannot decide what to answer till
 the end of word. We prove that automaton always has to branch at
 every $a_i$. Suppose contrary, there is $a_i$ where automaton goes
 to the same state whether it read $a_i = 0$ or $a_i = 1$. Then forward
 we give the next symbols $01^{n-2i}$ and automaton cannot decide what to
 answer. So it must branch for every $a_i$, we can say it $"remembers"$ this bit.
 But maybe it can merge ($"forget"$) afterwards? No, because
 constructions\par
\begin{figure}
  \centering
  \includegraphics[height=2.5cm]{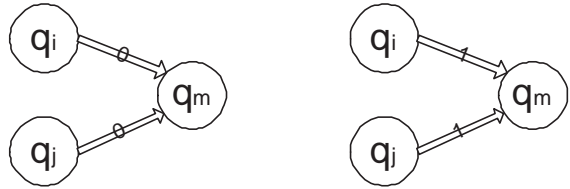}\\
  \caption{From different states go to one with the same input symbol}\label{viens}
\end{figure}
 are forbidden by reversibility, but construction\par
\begin{figure}
  \centering
  \includegraphics[height=2.5cm]{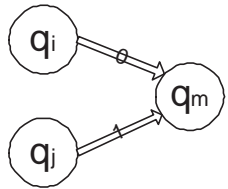}\\
  \caption{From different states go to one with different input symbols}\label{viens}
\end{figure}
 by the same reason as branching must occur (for all states $q_i$, $q_j$, $q_m$, i $\neq$ j).
 Then it follows that automaton $"remembers"$ all bits and the total number of states
 is at least $2^k$.
\par Second part: We use technique introduced by~\cite{a:n:t:v}. Let
 M be any n-restricted QFA accepting $L_n$ with probability $p > 1/2$.
 The following claim formalizes the intuition that the state of M after n symbols in
 form $a_11a_21a_31a_41a_51a_61...a_k1$ have been read is an
 $"encoding"$ (in case of RFA, so deterministic, we said $"remember"$) of the $\{a_i\}$.
 \begin{claim}\par
    There is a serial encoding of k bits into $\mathbb{C}^Q$, and
    hence into $\lceil \log|Q|\rceil$ qubits, where Q is the set
    of basis states of the M.
 \end{claim}
 {\it Proof.} Let $Q_{acc}$ and $Q_{rej}$ be the set of accepting and rejecting
 states respectively.  Let $U_\sigma$ be a unitary operator of M corresponding
 to the symbol $\sigma \in \{0, 1, $\o$, \$\}$.
 \par We define an encoding $f : \{0, 1\}^k \longrightarrow \mathbb{C}^Q$ of k-bit strings into
 unit superpositions over the basis states of the QFA M by letting $|f(x)\rangle$
 be the state of the automaton M after the input string
 $a_11a_21a_31a_41a_51a_61...a_k1$ where $a_i \in \{0, 1\}$ has
 been read. We assert that f is a serial encoding. \par
 To show that indeed f is such an encoding, we exhibit a suitable
 measurement for the $a_i$th bit for every $i \in [1..k]$. Let, for
 $y \in \{0, 1\}^{n - 2*i + 1}$, $V_i(y) =
 U_{\$}U^{n-2*i}_1U_0U^{-1}_{y_1}U^{-1}_{y_2}...U^{-1}_{y_{n-2*i-1}}U^{-1}_{y_{n-2*i}}U^{-1}_{y_{n-2*i+1}}$.
 The $i$th measurement then consists of first applying the unitary
 transformation $V_i(1a_{i+1}1...1a_k1)$ to $|f(x)\rangle$, and
 then measuring the resulting superposition with respect to $E_{acc}\otimes E_{rej}\otimes
 E_{non}$. Since for words with form $a_11a_21...1a_i01^{n-2*i}$, containment in $L_n$ is
 decided by the $a_i$, and because such
 words are accepted or rejected by then n-restricted QFA M with
 probability at least p only after the entire input has been read,
 the probability of observing $E_{acc}$ if $a_i = 0$, or $E_{rej}$ if $a_i =
 1$, is at least p. Thus, f defines a serial encoding.
 \qed
 Then it follows from Theorem~\ref{thm:encod} that
 $\lceil \log|Q|\rceil = \Omega(\frac{k}{\log k})$, but since $k =
 \frac{n}{2}$, we have $|Q| = 2^{\Omega(\frac{n}{\log n})}$. From
 Theorem~\ref{thm:rr} it follows that any quantum automaton
 that recognize $L_n$ also require $2^{\Omega(\frac{n}{\log n})}$
 states.
\par Third part: Easy.
\par Fourth part: The PFA Q in
Theorem ~\ref{thm:pd} has one rejecting ($q_{rej}$), one accepting
($q_{acc}$), one initial ($q_{ini}$) state and many non-halting
states $q_i$. We build PFA $Q^{'}$ recognizing language $L_n$ with
one rejecting ($q^{'}_{rej}$), one accepting($q^{'}_{acc}$), one
starting ($q^{'}_{ini}$) state and several non-halting states
$q^{'}_{i, 0}$, $q^{'}_{i, 1}$ and $q^{'}_{i, 2}$, where i is from
set of states' indexes from automaton Q. For every transition from
state $q_i$ to state $q_j$ with probability p for the input symbol
a (we denote this by f($q_i$, a, $q_j$, p)) there are $6$
transitions in $Q^{'}$ (we denote it by f'):
\par
    \begin{enumerate}
    \item f'($q^{'}_{i, 0}$, $1$, $q^{'}_{i, 0}$, p)
    \item f'($q^{'}_{i, 0}$, $0$, $q^{'}_{i, 1}$, p)
    \item f'($q^{'}_{i, 1}$, $1$, $q^{'}_{i, 0}$, p)
    \item f'($q^{'}_{i, 1}$, $0$, $q^{'}_{i, 2}$, p)
    \item f'($q^{'}_{i, 2}$, $1$, $q^{'}_{i, 2}$, p)
    \item f'($q^{'}_{i, 2}$, $0$, $q^{'}_{i, 2}$, p)
    \end{enumerate}
\par
For every transformation f($q_{ini}$, $\o$, $q_i$, p), there is a
transformation f'($q^{'}_{ini}$, $\o$, $q^{'}_{i, 0}$, p). For
every f($q_i$, a, $q_{rej}$, p) there is f'($q^{'}_{i, k}$, $x$,
$q^{'}_{rej}$, p) such that for all $k \in \{0, 1, 2\}$, $x \in
\{0, 1\}$, and for every f($q_i$, $\$$, $q_{rej}$, p) there is
f'($q^{'}_{i, k}$, $\$$, $q^{'}_{rej}$, p) for all $k \in \{0, 1,
2\}$, and for any f($q_i$, $\$$, $q_{acc}$, p) there are
f'($q^{'}_{i, 2}$, $\$$, $q^{'}_{acc}$, p), f'($q^{'}_{i, 0}$,
$\$$, $q^{'}_{rej}$, p), f'($q^{'}_{i, 1}$, $\$$, $q^{'}_{rej}$,
p).
\par
Informally, we make 3 copies from states in Q and their meaning is
similar than for states of automaton from Lemma ~\ref{lemma:div0}.
Automata computes parallel two things: is length of input word $n$
and is there any adjacent zeroes in it. It is obviously that the
accepted words are those whose length is $n$ and there are two
adjacent $0$ in them.\par\qed

\section{Conclusion}
\par We have shown that sometimes quantum automata must be almost
doubly exponential larger than classical automaton. But there
still remains open the other question. As follows from result of
Ambainis and Freivalds~\cite{a:f}, any language accepted by a QFA
with high enough probability can be accepted by a RFA which is at
most exponentially bigger that minimal DFA accepting the language.
Thus follows that Theorem~\ref{thm:pr} is close to maximal gap
between probabilistic and quantum automaton with high enough (this
was precisely computed by Ambainis and \c Kikusts~\cite{a:k} -
greater than $\frac{52+4\sqrt{7}}{81} = 0.7726...$) probability of
success. But it is not clear how it is when we allow smaller
probability of correctness. Author do not now any lower or upper
bound in this case.

\section*{Acknowledgements}
\par I would like to thank R\= usi\c n\v s Freivalds for suggesting
the problem and help during research.


\end{document}